\documentclass{osa-article}

\journal{oe}

\articletype{Research Article}

\usepackage[separate-uncertainty = true]{siunitx}
\usepackage{chemformula}
\usepackage{cancel}

\newcommand{\textadd}[1]{#1}
\newcommand{\textdel}[1]{}

\begin{document}

\title{Analysis of silicon nitride partial Euler waveguide bends}

\author{Florian~Vogelbacher,\authormark{1,2,*} Stefan~Nevlacsil,\authormark{1} Martin~Sagmeister,\authormark{3} Jochen Kraft,\authormark{3} Karl~Unterrainer,\authormark{2} and Rainer~Hainberger\authormark{1}}

\address{\authormark{1}AIT Austrian Institute of Technology GmbH, Center for Health \& Bioresources, Giefinggasse~4, 1210~Vienna, Austria\\
\authormark{2}TU Wien, Photonics Institute, Gusshausstra{\ss}e~27-29, 1040~Vienna, Austria\\
\authormark{3}ams~AG, Tobelbader~Stra{\ss}e~30, 8141~Premst{\"a}tten, Austria
}

\email{\authormark{*}florian.vogelbacher@tuwien.ac.at} 



\begin{abstract} 
In this work we present a detailed analysis of individual loss mechanisms in silicon nitride partial Euler bends at a wavelength of \SI{850}{\nano\meter}. This structure optimizes the transmission through small radii optical waveguide bends. The partial Euler bend geometry balances losses arising from the transition from the straight to the bend waveguide mode, and radiative losses of the bend waveguide mode. Numerical analyses are presented for 45-degree bends commonly employed in S-bend configurations to create lateral offsets, as well as 90- and 180-degree bends. Additionally, 90-degree partial Euler bends were fabricated on a silicon nitride photonic platform to experimentally complement the theoretical findings. The optimized waveguide bends allow for a reduced effective radius without increasing the total bend loss and, thus, enable a higher component density in photonic integrated circuits.
\end{abstract}

\section{Introduction}
Photonic integrated circuits (PICs) play an important role in data- and telecommunication \cite{Vlasov.2012,Cheng.2018} and are attracting increasing interest in other fields of applications such as astronomy \cite{BlandHawthorn.2009, Norris.2019}, biosensing \cite{Ciminelli.2019, Jans.2018}, medical diagnostics \cite{TalebiFard.2013, Hainberger.2019}, and quantum photonics \cite{Lenzini.2018}. These applications benefit from the strongly confined light, stable operation, and the small footprint compared to bulk optics. However, the compactness of a PIC is significantly limited by losses arising from waveguide bends used in routing of optical signals. Therefore, the optimization of bends is of pivotal importance in the design of high-density PICs or devices requiring ultra-low losses. 

Different approaches to reduce total bend losses have been proposed in literature. For example, a lateral offset between the incoming straight waveguide and the bend can reduce the mode mismatch between these two sections \cite{Kitoh.1995}. However, this technique relies on a high fabrication resolution, which is not always attainable by standard photolithography methods. Fabrication limitations are relaxed when a smooth transition between the straight waveguide and the bend is introduced. Ideally, such a transition allows to adiabatically convert the mode profile of a straight waveguide to that of a bent section. In the matched bend approach the length of the bend is optimized to minimize the higher order leaky modes at the output of the bend \cite{Melloni.2003}. Alternatively, a variation of the waveguide width, a multi-step patterning with grooves along the bend \textadd{or subwavelength gratings on the waveguide} can \textdel{increase} \textadd{alter} the mode confinement and reduce losses in the bend \cite{Harjanne.2004, Song.2016, Wu.2019}. The implementation of groves \textadd{and subwavelength gratings} requires additional fabrication steps and increases costs. In many applications a bend loss reduction by an optimized geometry alone is therefore favorable. Various bend geometries have been proposed, e.g.\ based on spline curves \cite{Harjanne.2004}, trigonometric functions \cite{Liu.1991}, or deduced from a topology optimization \cite{Iguchi.2018, Gabrielli.2012, Koos.2007, Liu.2019}. The variational method \cite{Chen.2012b, Bahadori.2019} relies on analytic bend loss models for deriving an optimized bend geometry. However, this approach is inherently limited by the physical validity of the applied bend loss model and, thus, carries the risk of omitting or inadequately describing critical loss mechanisms.

A bend geometry of particular interest is based on the Euler spiral, also known as a clothoid. Euler spirals are widely employed in civil engineering as transition curves because of the continuous change in curvature and straightforward implementation \cite{AmericanRailwayEngineering.2003, Kuhn.2013}. An Euler spiral is defined through a linear increase of the bend curvature along the path length, resulting in a continuous transition from the straight waveguide to the waveguide bend. This transition reduces the excitation of higher order modes \cite{Cherchi.2013}. Optical waveguide bends based on Euler spirals have been successfully demonstrated for single mode and multimode optical waveguides \cite{Cherchi.2013, Jiang.2018}. The combination of an Euler bend with a section of constant curvature has been presented by Fujisawa et al.\ for a silicon photonic platform in \cite{Fujisawa.2017}. This combination balances two major loss components of a waveguide bend, namely the transition losses arising from a changing curvature, and radiative losses inherent to a bend waveguide mode.

In this work, we present a detailed numerical analysis of the partial Euler bend geometry. A reformulated description to construct the partial Euler bend geometry compared to the work of Fujisawa et al. \cite{Fujisawa.2017} facilitates the implementation. This formulation allows to implement arbitrary angled bends $<\ang{360}$. Here, we optimize the partial Euler bend parameter for \ang{45}, \ang{90} and \ang{180} bends for a low-loss silicon nitride photonic platform \cite{Hainberger.2019, Pfeiffer.2018, Porcel.2019, Munoz.1272018212018} at a wavelength of \SI{850}{\nano\meter}. In difference to previous publications of other groups analyzing various kinds of waveguide bend geometries, our numerical analysis is based on the eigenmode expansion method, which allows a separation of the total losses into bend mode radiation and transition losses along the geometry. This separation provides a valuable insight into the individual contributions to the total loss. Bend angles larger than \ang{180}, which are used for instance in Sagnac loop mirrors \cite{Ren.2016, Zhang.2014}, can also be constructed with the presented formalism but are not addressed in this work. In addition to the in-depth numerical analysis, we present experimental results on the bend losses of partial Euler bends with an effective radius of $\SI{50}{\micro\meter}$ fabricated on a silicon nitride photonic platform.

\section{Theory}
\begin{figure}[htb]
\centering
\includegraphics{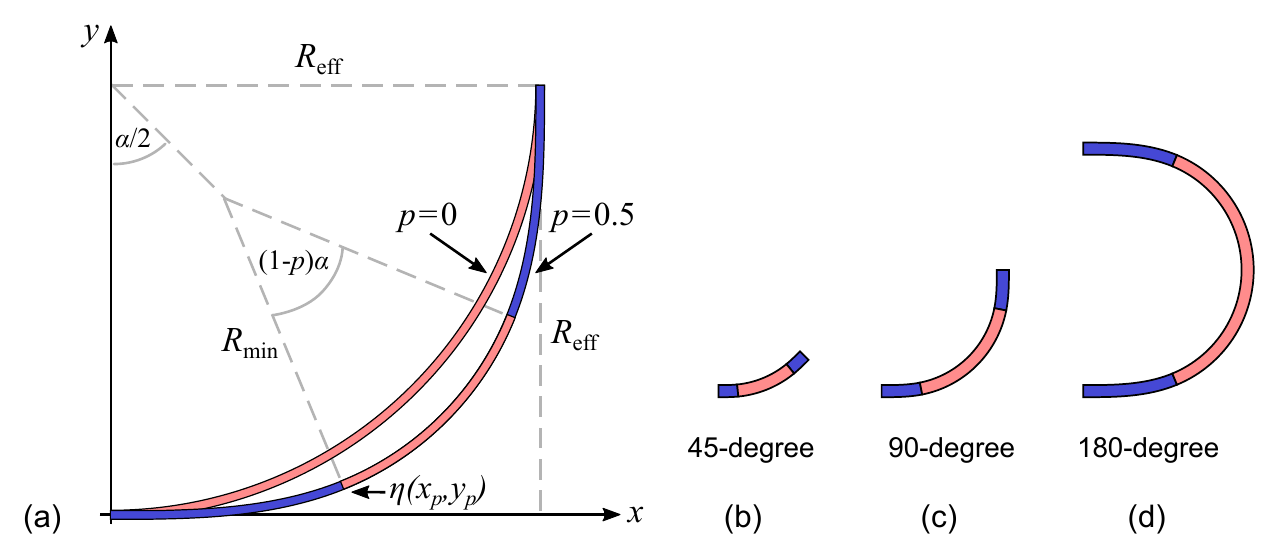}
\caption{(a)~Geometry of a \ang{90} partial Euler bend with bend parameter $p=0.5$ and a circular bend. The design parameter $p$ influences the angle at which the transition between the section of linearly increasing curvature (blue) to the section with constant curvature (light red) occurs. For $p=0$ the partial Euler bend is equivalent to a circular bend. The middle section has a radius of curvature of $R_\text{min}$. (b)~\ang{45}, (c)~\ang{90}, and (d)~\ang{180} bend geometries for $p=0.2$.}
\label{fig:vogel1}
\end{figure}

In this section, the geometric properties of an Euler spiral are presented and subsequently used for the construction of a partial Euler bend consisting of a section with linearly increasing curvature and a section of constant curvature. Figure~\ref{fig:vogel1}(a) exemplarily depicts the geometry of a \ang{90} partial Euler bend and indicates the parameters used to describe a partial Euler bend. The bend parameter $p$ relates to the portion of the bend having a linearly increasing curvature. The input and output of a partial Euler bend with effective radius $R_\text{eff}$ coincide with that of a circular bend of the same radius. The tangent lines of the input and output are preserved. The bend geometry is expressed in Cartesian coordinates to enable a straightforward implementation in a photonic design software.

\subsection{The Euler spiral}
In Cartesian coordinates the Euler spiral can be constructed through the scaled Fresnel integrals \cite{Olver.2010}
\begin{subequations}
\begin{align}
    \centering
    x(s) &= \int_0^s \cos{\left(\frac{t^2}{\textadd{2R_0^2}}\right)}\;dt \label{eqn:Fresnel_integral_a} \\
    y(s) &= \int_0^s \sin{\left(\frac{t^2}{\textadd{2R_0^2}}\right)}\;dt \label{eqn:Fresnel_integral_b}
\end{align}
\end{subequations}
where $s$ denotes the \textdel{normalized} path length \textadd{and $R_0$ is a free to choose radius. For example, this parameter can be set to $1/\sqrt{2}$ unit length for ease of calculation of the Fresnel integrals}. The curvature $\kappa$ increases linearly with the path length, i.e.,
 \begin{equation}
    \centering
    \kappa (s) = \frac{\textdel{2}s}{\textadd{R_0^2}}
    \label{eqn:kappa_path_length}
\end{equation}
and the angle $\alpha$ is reached at a path length
 \begin{equation}
    \centering
    s(\alpha) = \textadd{R_0}\sqrt{\textadd{2}\alpha}.
    \label{eqn:path_length}
\end{equation}
The curvature at an angle $\alpha$ is accordingly given by 
 \begin{equation}
    \centering
    \kappa(\alpha) = \textadd{\frac{\sqrt{2\alpha}}{R_0}}.
    \label{eqn:kappa_angle}
\end{equation}
The integrals in Eq.~\eqref{eqn:Fresnel_integral_a} and Eq.~\eqref{eqn:Fresnel_integral_b} cannot be solved analytically, but numerical integration or a series expansion can be used to solve these with suitable precision \cite{Bulirsch.1967, Press.2007, Olver.2010, Alazah.2014}.

\subsection{The partial Euler bend for arbitrary angles}
The presented partial Euler bend of total bend angle $\alpha$ is designed symmetrically around $\alpha/2$. In the following the first half of the bend is described. The first section up to an angle $p\alpha/2$ is an Euler spiral, followed by a section with constant radius of curvature up to the angle $\alpha/2$. The second half is readily obtained from a mirroring operation. To construct a partial Euler bend with an effective radius $R_\text{eff}$, the functions describing the bend need to be rescaled. The rescaling factor depends on the bend angle $\alpha$, the effective radius $R_\text{eff}$, and the bend parameter $p$.

In a first step, the coordinates $(x_p,y_p)$ of the end of the Euler section are calculated from Eq.~\eqref{eqn:Fresnel_integral_a} and Eq.~\eqref{eqn:Fresnel_integral_b}, i.e., $x_p=x(s_p)$ and $y_p=y(s_p)$, where $s_p=\textadd{R_0}\sqrt{p\alpha\textdel{/2}}$ denotes the path length to reach the angle $p\alpha/2$. The radius of curvature $R_p$ at $(x_p,y_p)$ is given from Eq.~\eqref{eqn:kappa_path_length}, and results in
 \begin{equation}
    \centering
    R_p = \frac{1}{\kappa(s_p)}=\frac{\textadd{R_0}}{\textdel{2}\sqrt{p\alpha\textdel{/2}}}.
    \label{eqn:radius_p}
\end{equation}
The offset to translate the start of the circular section to the end of the Euler section is calculated with
\begin{subequations}
\begin{align}
    \centering
    \Delta x &= x_p - R_p \sin{(p\alpha / 2)} \label{eqn:delta_x} \\
    \Delta y &= y_p - R_p \left[1-\cos{(p\alpha / 2)}\right] \label{eqn:delta_y}.
\end{align}
\end{subequations}
The total bend length is the sum of the path lengths of two Euler sections and the circular section with radius $R_p$ resulting in
\begin{equation}
    \centering
    s_0 = 2s_p + R_p \alpha(1-p).
    \label{eqn:total_length}
\end{equation}
Combining the results, the functions describing the first half of the unscaled partial Euler bend for angles smaller than $\alpha/2$ are
\begin{subequations}
\begin{align}
    \centering
    x_\text{bend}(s) &=
        \begin{cases}
            x(s) & \text{for } 0 \leq s \leq s_p\\
            R_p \sin{\left( \frac{s-s_p}{R_p}+\frac{p\alpha}{2} \right)} + \Delta x    & \text{for } s_p < s \leq s_0/2 
        \end{cases}  \label{eqn:unscaled_euler_x}
    \intertext{and}
    y_\text{bend}(s) &=
        \begin{cases}
            y(s) & \text{for } 0 \leq s \leq s_p\\
            R_p \left[1 - \cos{\left( \frac{s-s_p}{R_p}+\frac{p\alpha}{2} \right)} \right] + \Delta y    & \text{for } s_p < s \leq s_0/2 
        \end{cases} \label{eqn:unscaled_euler_y}
\end{align}
\end{subequations}
From geometric considerations we find the rescaling factor $\eta$ to achieve an effective radius $R_\text{eff}$ to be
\begin{equation}
    \centering
    \eta = \frac{R_\text{eff}}{y_\text{bend}\left( s_0/2 \right) + x_\text{bend}\left( s_0/2 \right)/\tan{\left(\alpha/2 \right)} }.
    \label{eqn:rescale_factor}
\end{equation}
The equations describing the rescaled bend can be written as 
\begin{subequations}
\begin{align}
    \centering
    \tilde{x}_\text{bend}(s) &=  \eta\cdot x_\text{bend}{\left( s/\eta \right)} \label{eqn:rescaled_euler_x} \\
    \tilde{y}_\text{bend}(s) &= \eta\cdot y_\text{bend}{\left( s/\eta \right)}  \label{eqn:rescaled_euler_y}.
\end{align}
\end{subequations}
with the minimum radius of curvature given by
\begin{equation}
    \centering
    R_\text{min} = \eta\cdot R_p.
    \label{eqn:minimum_radius_of_curvature}
\end{equation}
The path length to construct the first half of the rescaled bend ranges from 
\begin{equation}
    \centering
    0 \leq s \leq \eta \frac{s_0}{2} .
    \label{eqn:path_length_range}
\end{equation}

The geometric construction allows for bend angles $<\ang{360}$. However, the denominator of the rescale factor $\eta$ in Eq.~\eqref{eqn:rescale_factor} depends on the parameter $p$ of the partial Euler bend as well as on the bend angle $\alpha$. For a bend angle of approximately \ang{263.3} and $p=1$ the denominator vanishes. In this case, no solution for a pure Euler bend exists but partial Euler bends with a circular section are still possible. For larger bend angles, the maximal possible bend parameter continuously decreases. For example, for a \ang{270} bend the maximum value of the bend parameter is $p=0.585$. Close to such a singularity in Eq.~\eqref{eqn:rescale_factor} the arc length and footprint of the partial Euler bend increases significantly. For the studied bends with $\alpha\leq \ang{180}$ no singularity occurs and the minimum radius of curvature $R_\text{min}$ is always smaller than the effective radius $R_\text{eff}$ for all $0< p\leq 1$.

\begin{figure}[hbt]
\centering
\includegraphics{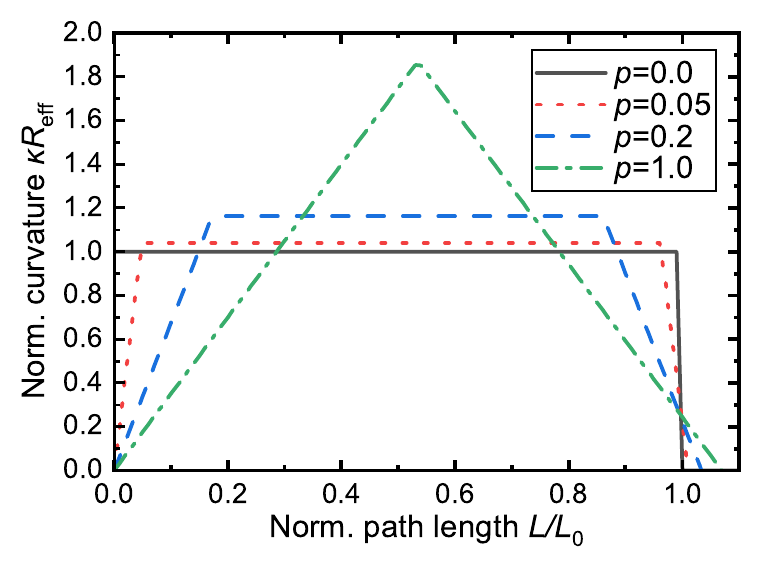}
\caption{Normalized curvature along a \ang{90} bend depending on the value of parameter $p$. The path length \textdel{was} \textadd{$L$ is} normalized to the arc length \textadd{$L_0$} of a circular bend\textdel{ of radius $R_\text{eff}$}. A pure Euler bend ($p=1$) shows an \SI{82}{\percent} increased curvature compared to a circular bend ($p=0$), which increases radiative bend mode losses. The normalized path lengths of partial Euler bends with $p>0$ are larger than unity.}
\label{fig:vogel2}
\end{figure}

Figure~\ref{fig:vogel2} plots the normalized curvature $\kappa R_\text{eff}$ profiles for four different values of the bend parameter $p$ for a \ang{90} bend, i.e.\ $p=0$, $0.05$, $0.2$, and $1$. The path length is normalized to the arc length \textadd{$L_0=R_\text{eff}\pi/2$} of a circular bend. The total path length does not change significantly for different \textadd{values of the} parameter $p$. The first derivative of the curvature is not continuous at the transition point between the Euler and the circular section. This \textadd{singularity} does not introduce additional losses because the curvature is continuous and, thus, the mode profile does not change at this transition. The curvature in the middle of a pure Euler bend ($p=1$) is \SI{82}{\percent} higher than that of a circular bend with constant radius. This causes increased radiative bend mode losses compared to a circular bend. On the other hand, for a circular bend the abrupt change in curvature at the interface between the straight and the bent section induces a mode mismatch. A balance between the two extreme geometries optimizes the total loss. 

\section{Numerical analysis}
\subsection{Simulation setup and parameters}
The modal power and the bend mode radiation in the partial Euler bend geometries were numerically determined with the commercial eigenmode expansion (EME) method tool FIMMPROP (PhotonD, FIMMWAVE) \cite{Gallagher.2003}. \textadd{The software employs a fully vectorial complex finite difference mode solver that allows the calculations of eigenmodes in a bent waveguide section. The imaginary part of the propagation constant of such an eigenmode reflects its radiative loss.}
The Euler section with linearly increasing/decreasing curvature was divided into $64$~bend subsections of equal path length $\Delta s$. A single section was used for the circular bend section. The symmetry of the Euler bend at $\alpha/2$ was exploited to reduce the computation time since only eigenmodes of the first half of the partial Euler bend were calculated. Straight sections at the beginning and end of the partial Euler bend were used to include the losses from the mode mismatch between the straight waveguide input/output and the bend. The transition loss arising from the gradual change of curvature is derived from the difference between total loss and radiative bend mode loss. The EME method is highly suitable for bend radii, at which the computational domain for FDTD would become too large to efficiently been solved.

The investigated wavelength was \SI{850}{\nano\meter} and the refractive indices of silicon nitride and silicon dioxide were assumed to be $2.02$ and $1.46$, respectively \cite{Luke.2015, RodriguezdeMarcos.2016}. For this generic silicon nitride photonic platform absorption, scattering, and substrate leakage are neglected because advanced fabrication technologies allow for $<\SI{1}{\decibel\per\centi\meter}$ propagation losses in the visible and near-infrared wavelength region \cite{Hainberger.2019, Pfeiffer.2018, Porcel.2019, Munoz.1272018212018}. Therefore, transition and radiative bend mode losses dominate for bends with small radii. The total path length changes only slightly for different values of the bend parameter $p$ with the same effective radius $R_\text{eff}$ and $\alpha < \ang{180}$, such that the general findings of the optimization hold also true for elevated propagation losses. 
For loss mechanisms such as strong absorption and scattering, which might be critical for other photonic integrated platforms, supplementary terms can be included in the employed numerical analysis approach.
The cross section of the embedded waveguide is set to \SI{700x160}{\nano\meter}, which ensures single mode operation. The influence of the bend geometry on the loss is investigated for the more sensitive TM-like polarized guided mode. Bend loss for TE-like polarization are in general smaller compared to the TM-like polarization because the confinement of the TE-like mode is higher and radiative losses are reduced. Therefore, in PIC designs where both polarizations are present in the optical path the TM-like polarization typically dictates the minimum usable effective bend radius. Figures~\ref{fig:vogel1}(b)-\ref{fig:vogel1}(d) depict the investigated \ang{45}, \ang{90}, and \ang{180} partial Euler bend geometries for the bend parameter value $p=0.2$. 

\begin{figure}[htbp]
\centering
\includegraphics{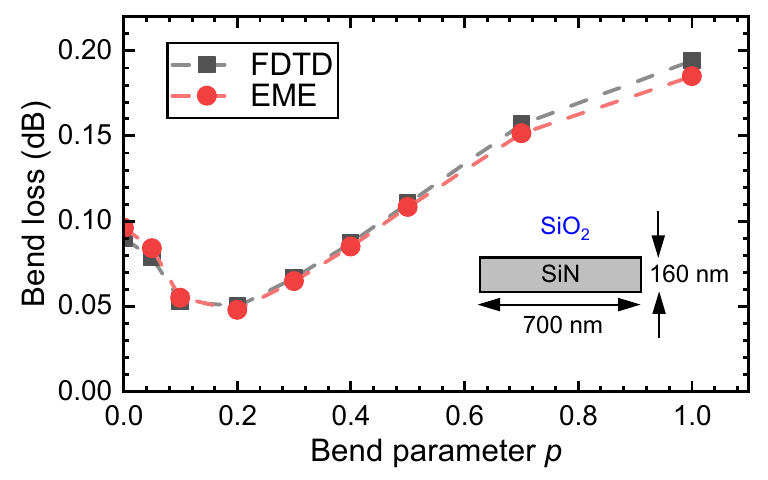}
\caption{Total bend loss determined by eigenmode expansion (EME) and finite-difference-time-domain (FDTD) method for \ang{90} bends, \SI{40}{\micro\meter} effective radius, and TM-like polarization.}
\label{fig:vogel4}
\end{figure}

The EME method was benchmarked against finite-difference-time-domain (FDTD) calculations (Lumerical) for \ang{90} bends with an effective radius of \SI{40}{\micro\meter}. Figure~\ref{fig:vogel4} compares the bend losses calculated by both FDTD and EME for different bend parameter $p$. The results agree within \SI{0.01}{\decibel}, or \SI{7}{\percent}, respectively. This confirms that the employed EME method is a suitable alternative to FDTD allowing computationally efficient calculations of waveguide bend losses.

\subsection{Results and discussion of numerical analysis}
\begin{figure}[htbp]
\centering
\includegraphics{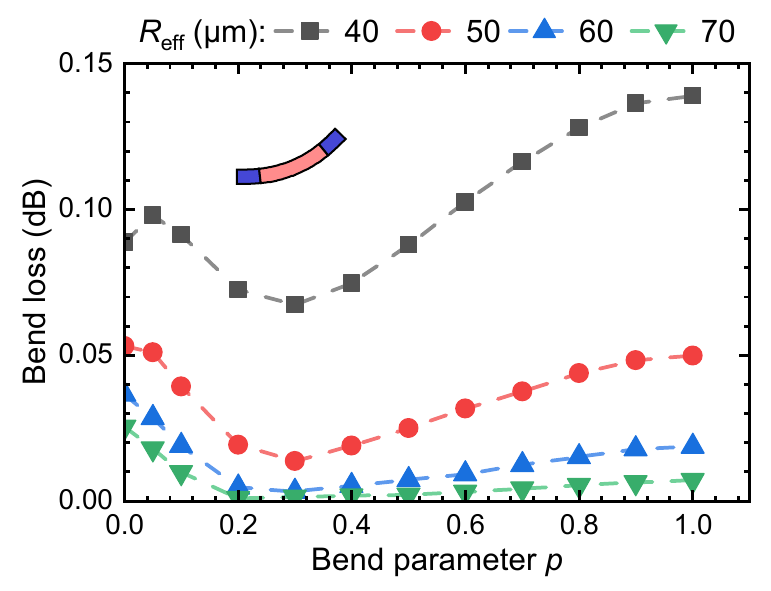}
\caption{Total bend loss through a \ang{45} partial Euler bend for different bend radii, TM-like polarization. Inset depicts the bend geometry.}
\label{fig:vogel5}
\end{figure}

Figure~\ref{fig:vogel5} summarizes the calculated total bend loss for the guided mode propagation through a \ang{45} partial Euler bend with varying bend parameter $p$. A bend parameter of $p=0.3$ provides optimal performance for the studied radii. For instance, the bend loss for $R_\text{eff}=\SI{40}{\micro\meter}$ can be reduced to \SI{0.07}{\decibel} ($p=0.3$) compared to \SI{0.09}{\decibel} ($p=0$) and \SI{0.14}{\decibel} ($p=1$). The loss through an S-bend geometry, consisting of a second rotated congruent bend, can be readily obtained by squaring the transfer function or doubling the loss of a single bend, respectively.

\begin{figure}[htb]
\centering
\includegraphics{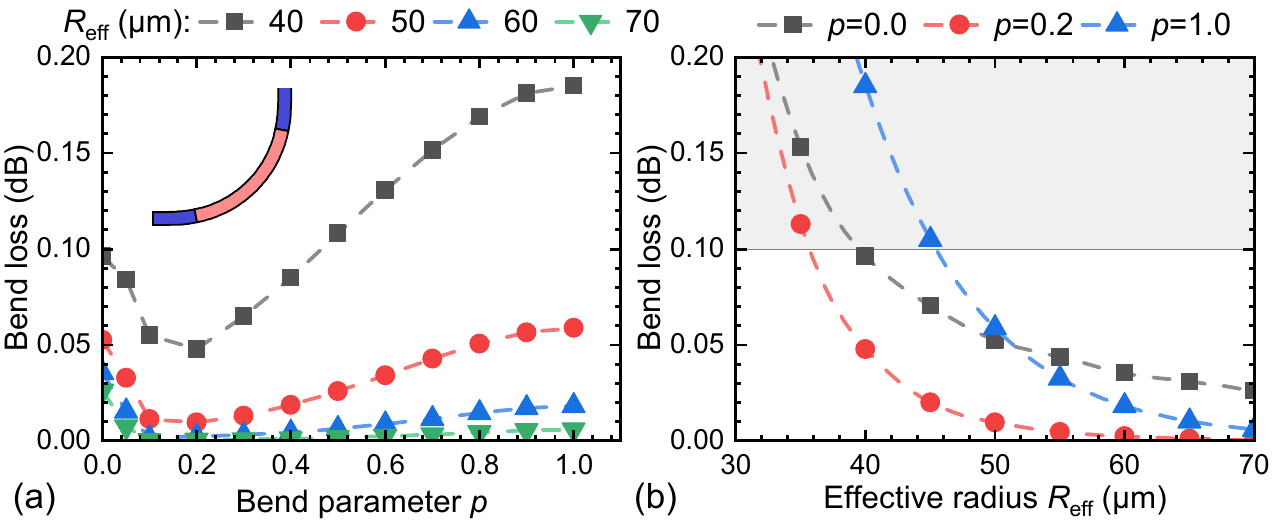}
\caption{Total bend loss through a \ang{90} partial Euler bend, TM-like polarization. (a)~Variation of bend parameter $p$ for different effective bend radii. (b)~Variation of effective bend radius $R_\text{eff}$ for different values of the bend parameter $p$. The area with loss larger than \SI{0.1}{\decibel} is indicated in grey for visualization.}
\label{fig:vogel6}
\end{figure}

Figure~\ref{fig:vogel6}(a) summarizes the numerical results of the calculated bend losses through \ang{90} bends. The results show a minor influence of the bend parameter $p>0.2$ for effective bend radii $R_\text{eff}$ larger than approximately \SI{70}{\micro\meter}. However, for $p<0.2$ the transition from the straight to the bent waveguide section is no longer adiabatic and the total bend loss increases from \SI{0.001}{\decibel} ($p=0.2$) for an optimized bend to \SI{0.026}{\decibel} ($p=0$) for a purely circular bend. At $R_\text{eff}=\SI{40}{\micro\meter}$ the total bend loss is \SI{0.05}{\decibel} for $p=0.2$, compared to \SI{0.10}{\decibel} for a circular bend ($p=0$), and \SI{0.19}{\decibel} for a pure Euler bend ($p=1$). One can notice that the loss through two \ang{45} bends in a S-bend structure will be larger than through a single \ang{90} bend for the same effective bend radius as the number of transitions from straight to bend waveguide sections is doubled in an S-bend structure.

Figure~\ref{fig:vogel6}(b) summarizes the total bend loss of a circular bend, an optimized Euler bend ($p=0.2$), and a pure Euler bend ($p=1$) for a varying effective radius $R_\text{eff}$. The optimized Euler bend provides the smallest losses over the studied range of effective bend radii. One can observe that for bend radii larger than \SI{50}{\micro\meter} a pure Euler bend is beneficial compared to a circular bend. However, below this radius, the reduced transition losses compared to the circular bend no longer compensate for the increasing radiative bend mode losses from the larger curvature of the pure Euler bend, resulting in a better performance of the circular bend compared to a pure Euler bend. For a targeted loss value of \SI{0.1}{\decibel} per bend, the effective bend radii are found to be \SI{36}{\micro\meter} ($p=0.2$), \SI{39}{\micro\meter} ($p=0$), and \SI{45}{\micro\meter} ($p=1$), respectively. Therefore, an optimized partial Euler bend can increase the component integration density for the given example by a factor of 1.17 relative to a circular bend. For comparison, if we choose the targeted loss value based on the loss of a \SI{70}{\micro\meter} circular bend of \SI{0.026}{\decibel} per bend, the partial Euler bend with the same loss has an effective radius of \SI{43}{\micro\meter} ($p=0.2$) and \SI{57}{\micro\meter} ($p=1$), respectively, leading to an improvement in integration density by a factor of $2.7$ compared to a circular bend, and a factor of $1.8$ compared to a pure Euler bend.

\begin{figure}[hbt]
\centering
\includegraphics{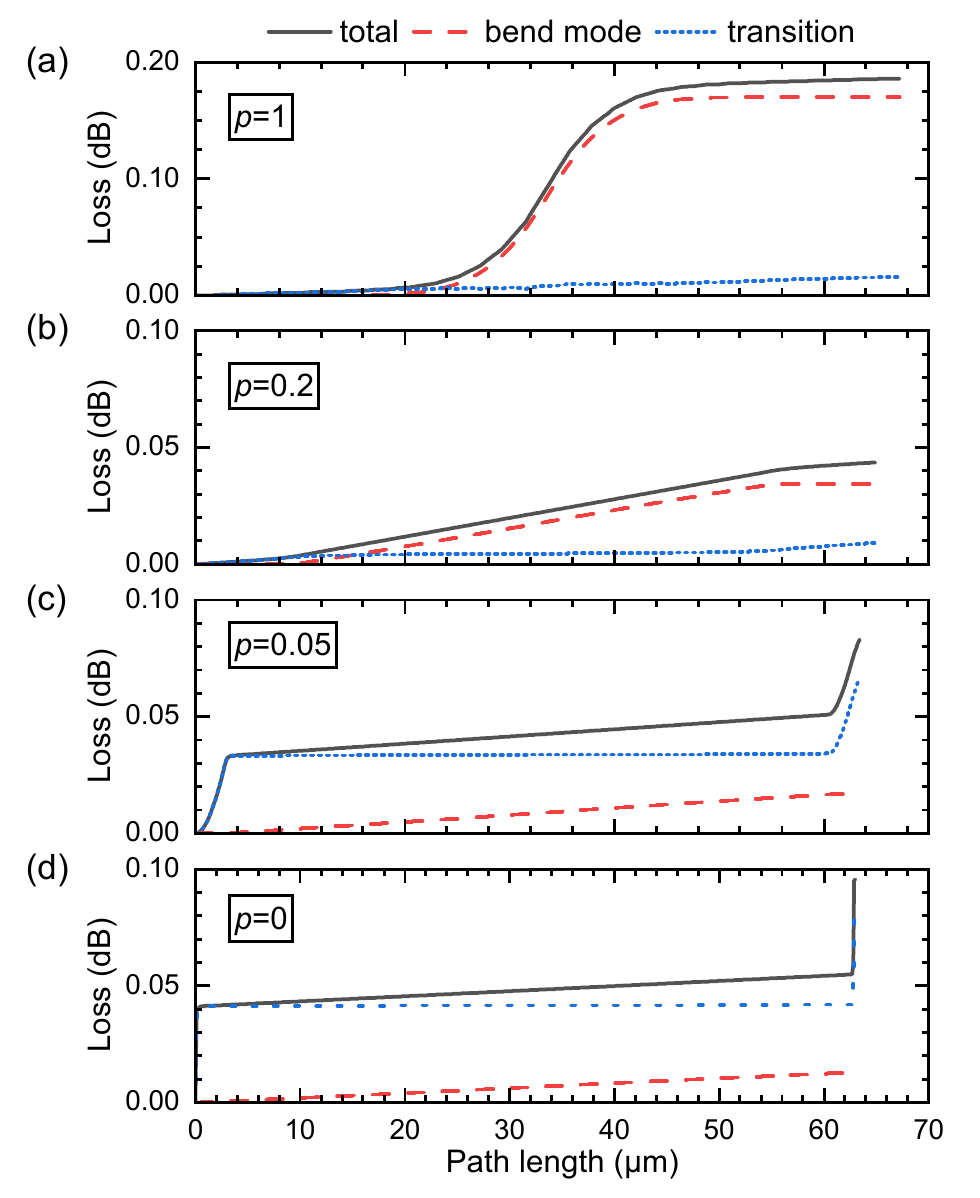}
\caption{Bend mode radiative, transition, and total losses accumulated along \ang{90} partial Euler bends with effective radius of \SI{40}{\micro\meter} for (a)~$p=1$, (b)~$p=0.2$, (c)~$p=0.05$, and (d)~$p=0$. The scale in (a) for the loss has twice the range than (b)-(d) for better visualization. An optimized partial Euler bend ($p=0.2$) distributes the accumulated loss along the bend. In contrast, a pure Euler bend ($p=1$) suffers from high radiative losses in the middle of the bend where curvatures are large. The geometry with $p=0.05$ is close to a circular bend ($p=0$) and the transition losses in the Euler sections are the major loss.}
\label{fig:vogel7}
\end{figure}

Figure~\ref{fig:vogel7}(a) to Fig.~\ref{fig:vogel7}(d) depict the individual contributions of the two loss mechanisms, i.e., transition and radiative bend mode losses, to the total bend loss along the bend for an effective radius of $R_\text{eff}=\SI{40}{\micro\meter}$ and four distinct values of the bend parameter ($p=1$, $p=0.2$, $p=0.05$, $p=0$). The optimized partial Euler bend ($p=0.2$) balances radiative bend mode losses arising from sections of high curvature with transition losses from sections of rapidly changing curvature.

Transition losses are not limited to the case of a discontinuous jump in curvature, which occurs for example at the interface between the straight input and bend section of a circular bend, but are present whenever a change in curvature $d\kappa/ds$ is present. In the circular middle section of a partial Euler bend, no transition losses occur ($d\kappa/ds=0$) and only the radiative bend mode components contribute to the loss. In contrast to the optimized bend, the pure Euler bend ($p=1$) exhibits large radiative bend mode losses at the middle of the bend originating from the increased curvature compared to a circular or partial Euler bend leading to significant total losses. The bend geometry with $p=0.05$ is close to a circular bend and has high losses at the start and end of the bend, where the transition loss arising from the fast changing curvature contribute most to the total loss. On the other hand, radiative bend mode losses are small for this geometry, as the minimal radius of curvature is larger than the minimal radius of curvature of the optimized and pure Euler bend.

\begin{figure}[htb]
\centering
\includegraphics{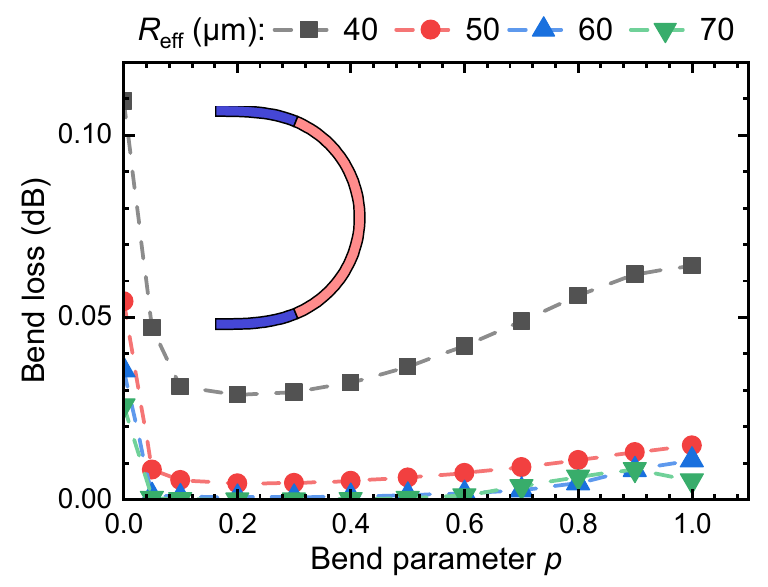}
\caption{Total bend loss through \ang{180} Euler bends, TM-like polarization. Inset depicts the bend geometry.}
\label{fig:vogel9}
\end{figure}

Like the \ang{90} bend, the \ang{180} bend shows lowest loss for a partial Euler bend with the parameter $p$ around $0.2$. The bend loss decreases from \SI{0.083}{\decibel} to \SI{0.047}{\decibel} for an effective bend radius of $R_\text{eff}=\SI{40}{\micro\meter}$. Figure~\ref{fig:vogel9} shows the bend loss for effective bend radii ranging from \SI{40}{\micro\meter} to \SI{70}{\micro\meter}. In contrast to \ang{45} and \ang{90} bends, the minimum radius of curvature $R_\text{min}$ of a \ang{180} bend is close to a circular bend for all $p$. This results in a reduced dependence of the bend loss on the partial Euler bend parameter for $p>0.2$.

In summary, the findings underline that the implementation of a pure Euler bend and other classes of bend geometries with sections of small radius of curvature can introduce pronounced radiative bend mode losses. Compared to a pure Euler bend a partial Euler bend increases the minimum radius of curvature $R_\text{min}$ at the expense of transition losses. This trade-off reduces the total bend loss. It has to be pointed out that the optimized values for the Euler bend parameter $p$ are specific to the investigated waveguide system. The waveguide geometry, the material refractive indices, the wavelength, and the polarization determine the confinement of the guided mode and, thus, strongly influence the radiation properties of curved waveguides. Therefore, a dedicated optimization for a specific waveguide system, effective bend radius, and polarization is mandatory. For a \ang{90} partial Euler bend on silicon photonic platform Fujisawa et al.\ obtained an optimal ratio of \SI{68}{\percent} between the Euler section length and the total bend length with a radius of $R_\text{eff}=\SI{3}{\micro\meter}$ for the TE-like polarized mode at a wavelength of \SI{1550}{\micro\meter} \cite{Fujisawa.2017}. This value corresponds to $p=0.52$ in our parameterization, which is considerably larger than the value of $p=0.2$ at $R_\text{eff}=\SI{40}{\micro\meter}$ in our study for a silicon nitride photonic platform. 

\section{Experimental analysis}
\subsection{Fabrication \textadd{and optical characterization}}
\begin{figure}[htb]
\centering
\includegraphics{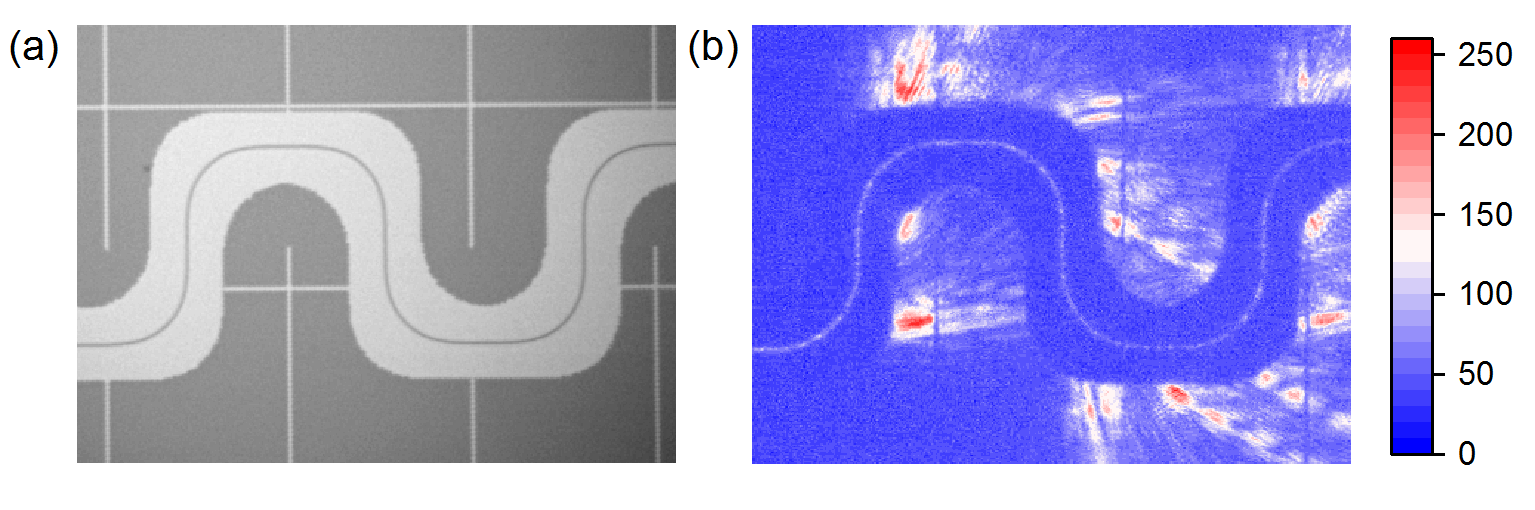}
\caption{(a)~Microscope image of fabricated \ch{SiN} partial Euler bends with an effective radius \SI{50}{\micro\meter}. (b)~Near-infrared image of radiation emitted from the bends and scattered at the lithography fill pattern.}
\label{fig:vogel10}
\end{figure}

Partial Euler bends of \ang{90} with an effective radius of \SI{50}{\micro\meter} and different values of the bend parameter $p$ have been fabricated on a \ch{SiN} photonic platform \cite{Sagmeister.2018}. The \ch{SiN} waveguide layer has a \textadd{nominal} thickness of \SI{160}{\nano\meter}. Figure~\ref{fig:vogel10}(a) shows a microscope image of these structures. The width of the waveguide was measured by scanning electron microscopy to be \SI{680}{\nano\meter}. The waveguide cross-section ensures single mode operation. Refractive indices of the \ch{SiO2} cladding and the \ch{SiN} waveguide material have been determined by ellipsometry to be $n_\text{SiO2}=1.46$ and $n_\text{SiN}=1.92$ at a wavelength of \SI{850}{\nano\meter}. The test structures consist of $240$ cascaded \ang{90} bends for the TE-like polarization. Each bend is separated from the subsequent bend by a \SI{10}{\micro\meter} straight segment. To account for the high bend losses for the TM-like polarization, a cleaved sample with four bends for the TM-like polarization was prepared. 

\textadd{To characterize the bend losses light from a \SI{850}{\nano\meter} Ti:sapphire laser source was end-face coupled via a polarization maintaining fiber (NA~$0.12$) to the waveguides with the bend structures. The polarization plane of the fiber was adjusted with a mechanical rotator and verified using a linear polarizer in combination with an optical power meter to couple to the TE-like and TM-like PIC waveguide mode, respectively. The output light of the PIC was collected with a single mode fiber (NA~$0.12$) connected to an optical power meter. The positioning of the input and output fibers to the PIC was supported by a piezoelectric auto alignment system (Thorlabs, NanoTrak\textregistered). The repeatability of the transmission power level measurement was better than \SI{0.2}{\decibel}.} A reference \textadd{waveguide} without bends was used to account for the \textdel{coupling} efficiency of the end-face fiber coupling into and out of the PIC, and the propagation loss from the routing of the optical signals. \textadd{The difference of transmitted optical power between the waveguide with the bend structures and the reference waveguide is divided by the number of bends, i.e.\ $240$ for the TE-like and $4$ for the TM-like mode, to calculate the loss of a single bend.} The numerical simulations include propagation losses of \SI{0.7}{\decibel} for the TE-like and \SI{0.4}{\decibel} for the TM-like mode \cite{Hainberger.2019}.

\subsection{Results and discussion of experimental analysis}
The radiation arising from the transition losses and bend waveguide modes for the TM-like polarization was recorded with an infrared camera (Hamamatsu, IR Vidicon C2741) and is visualized in Fig.~\ref{fig:vogel10}(b). The light enters the bend structure from the left handside. Strong radiation patterns are visible at each bend. Tracing the streaks back to the source suggests the main contribution from the middle of the bend, where the curvature is largest. 

\begin{figure}[htb]
\centering
\includegraphics{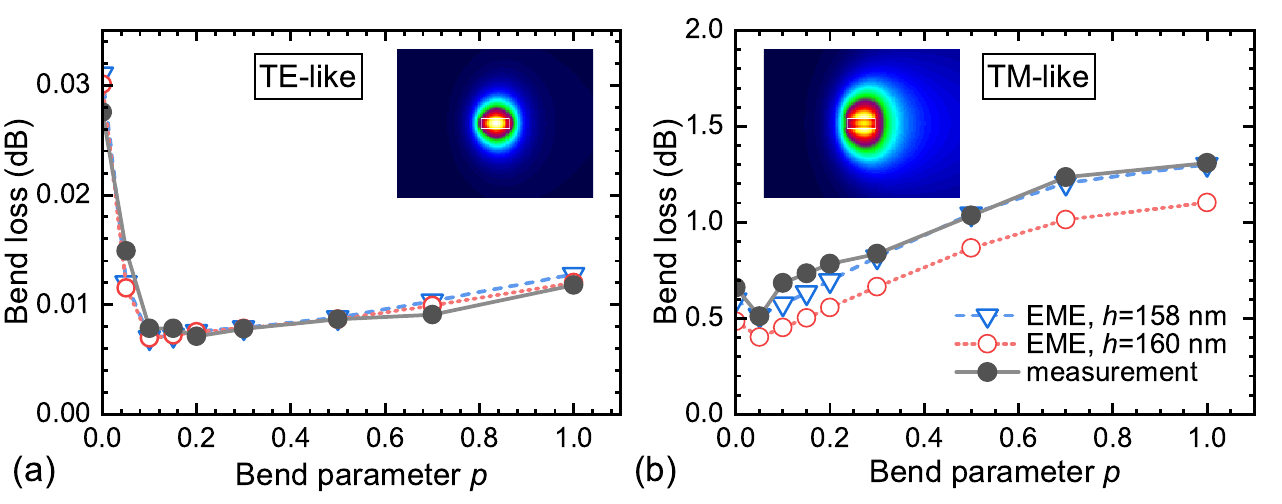}
\caption{Total bend loss measurement and numerical results through \ang{90} partial Euler bends with effective radius of \SI{50}{\micro\meter} for (a)~TE-like and (b)~TM-like waveguide mode. Insets show the bend waveguide profile for the major electric field components at the smallest radius of curvature of an Euler bend.}
\label{fig:vogel11}
\end{figure}

Figures~\ref{fig:vogel11}(a) and \ref{fig:vogel11}(b) summarize the influence of the bend parameter $p$ on the total bend losses in TE-like and TM-like polarization, respectively. For the TE-like polarization the mode mismatch between the straight and bend section is a major loss component and the largest bend losses occur, hence, for a circular bend with \SI{0.028}{\decibel} per bend. With an optimized partial Euler bend, the bend loss is reduced to \SI{0.007}{\decibel} at $p=0.2$. For larger values of the bend parameter $p$, the losses start to slowly increase. The numerical simulations with the EME method reflect the measurements to a high degree. For the TM-like waveguide mode the major losses arise from the radiative bend modes. Therefore, the optimal bend geometry converges to a circular bend at which the radius of curvature is maximized. A pure Euler bend ($p=1$) is detrimental to the bend performance. The offset between experiment and numerical analysis for the TM-like polarization can most likely be attributed to the effective thin film thickness of the SiN waveguide layer, as the bend losses of the TM-like mode are highly sensitive to variations of the waveguide geometry. \textadd{For example, a small change of the waveguide height in the simulations from \SI{160}{\nano\meter} to \SI{158}{\nano\meter} increases the bend loss of the TM-like mode by $\SI{0.2}{\decibel}$ for a pure Euler bend. This results in an improved agreement with the measured values. For the TE-like mode, the simulation results indicate a much lower influence of the waveguide height on the bend losses. Overall,} the TM-like mode shows increased bend losses of up to two magnitudes higher than the TE-like polarization for the chosen bend radius, which underlines the importance to optimize the TM-like mode in mixed polarization designs.

The SiN of the employed PIC technology platform has a refractive index which is \SI{5}{\percent} smaller than the value taken from literature and used in the numerical analysis in the previous section. This moderate difference in refractive index causes a strong increase of the minimum bend loss for the \SI{50}{\micro\meter} bends in the TM-like polarization from \SI{0.01}{\decibel} to \SI{0.5}{\decibel} due to the reduced confinement strength of the waveguide mode. Furthermore, the optimum value of the bend parameter found for the fabricated bends shifts towards zero, i.e.\ a circular bend, because the bend mode losses play a more important role than for the mode in the theoretical analysis with the higher SiN refractive index. This is in accordance with the prediction made by the EME calculations.


\section{Conclusion}
In this work the major loss mechanisms for partial Euler bend geometries were numerically analyzed with the EME method. This method provides additional insights, not accessible with other tools, into the individual contributions to the total loss along the geometry. Optimal values for the bend parameter have been identified for \ang{45}, \ang{90} and \ang{180} silicon nitride waveguide partial Euler bends operated with TM-polarization using literature values for the refractive indices at an operation wavelength of \SI{850}{\nano\meter}. The proper choice of the bend parameter balances transition losses arising from the continuous transition between sections of changing curvature, and radiative losses originating from the curvature of the bend waveguide mode. Transition losses are not limited to the abrupt change between a straight and a bent section, but occur whenever the curvature is changing non-adiabatically, e.g.\ for all bends with a significantly fast change in curvature. Compared to a pure Euler bend without a section of constant curvature, an optimized partial Euler bend exhibits a reduced bend loss. With a partial Euler waveguide bend geometry the integration density of PICs can be increased and losses for complex designs relying on a large number of bends reduced.  The numerical analysis has been applied to fabricated partial Euler bends for a specific silicon nitride photonic platform. The measurements of the bend losses agree well with the EME simulations and confirm the feasibility of this method to optimize waveguide bends for a specific material system and waveguide geometry. The small differences between the refractive indices assumed in the theoretical study and the experimental photonic platform showed a considerable impact on the optimal value of the bend parameter and on the actual loss values. Our findings emphasize that the optimal ratio between the Euler section to the total bend length has to be determined individually for different waveguide systems. 

\section*{Funding}
Austrian Research Promotion Agency (FFG) (850649); \textadd{European Union's Horizon 2020 research and innovation programme (688173)}.

\section*{Acknowledgments}
This research has received funding through the grant PASSION (No.~850649) from the Austrian Research Promotion Agency (FFG) \textadd{and European Union's Horizon 2020 research and innovation programme under the grant OCTCHIP (No.~688173)}.

\section*{Disclosures}
The authors declare no conflicts of interest.




%

\end{document}